\documentstyle[sprocl,epsfig]{article}

\def\Journal#1#2#3#4{{#1} {\bf #2}, #3 (#4)}

\def\NPB{{\em Nucl. Phys.} B}
\def\PLB{{\em Phys. Lett.}  B}
\def\PRL{\em Phys. Rev. Lett.}
\def\PRD{{\em Phys. Rev.} D}
\def\PRC{{\em Phys. Rev.} C}

\def\PRL{\em Phys. Rev. Lett.}
\def\PRE{\em Phys. Rept.}

\def\be{\begin{equation}}
\def\ee{\end{equation}}
\def\bea{\begin{eqnarray}}
\def\eea{\end{eqnarray}}
\begin{document}

%***************** start title ******************************************
\title{Simulation of SU(2) Dynamical Fermion at Finite Chemical Potential
and at Finite Temperature\footnote{Talk presented by Y.Liu\\
E-mail: liu@hirohe.hepl.hiroshima-u.ac.jp
}
}
%*****************end title *******************************************

%****************start author ****************************************
\vspace{10mm}

\vspace{20mm}
\author{Y.Liu$^{1}$
,O.Miyamura$^{1}$,
A.Nakamura$^{1}$,T.Takaishi$^{2}$}

%***************end author *************************************

%***************start address*************************************

\address{
$^{1}$Hiroshima University,Higashi--Hiroshima 739--8521, Japan\\
$^{2}$Hiroshima University of Economics,Hiroshima 731--01 Japan}

%**************end address *************************************


%***********start abstracts**********************************************
\maketitle

\abstracts{
SU(2) lattice gauge theory with dynamical  fermion at non-zero
chemical potential and at finite temperature is studied.
We focus on the influence of chemical potential for quark 
 condensate and mass of
pseudoscalar meson at finite temperature.
 Hybrid  Monte Carlo  simulations  with $N_f=8$ staggered  fermions
 are carried out on $12 \times 12\times 24 \times 4$ lattice.
At $\beta=1.1$ and $m_{q}=$0.05,0.07,0.1,
we calculate the quark condensate  and  masses of pseudoscalar meson
consisting of light and heavier quarks for chemical potential
$\mu=$ 0.0,0.02,0.05,0.1,0.2.
}
%**************end abstracts********************************************

%***********************start  section 1 *********************
\section{INTRODUCTION}
The study of strong interactions at finite baryon density has a long
history.
Recent theoretical speculation is that the ground state of dense baryonic
 matter may be more exotic~\cite{al}$^{,}$~\cite{ra}.
 Nambu-Jona-
Lasinio type and instanton liquid models~\cite{ra}$^{,}$~\cite{ba}
suggest that the QCD vacuum at sufficiently high baryon density could
become a color superconductor~\cite{ba}.

 Considerable progress has been made in simulating lattice quantum
chromodynamics ( LQCD ) at nonzero temperature~\cite{ka}.
But, there has been little progress in understanding the theory
at nonzero chemical potential even though a pioneering work
was presented sixteen years ago~\cite{na}.
The basic difficulty in simulation of LQCD at finite density
 is that the effective action becomes complex due to the introduction
of chemical potential.
  Standard algorithms such as hybrid molecular dynamics ( HMD )~\cite{du}
or hybrid Monte Carlo ( HMC )~\cite{dk} can not
be applied in this situation. In order to avoid the complex determinant,
 early simulations used quenched approximation,
but they do not correspond to the real world~\cite{da}.
In this situation, a possible way is to examine response of physical
quantities
to chemical potential at $\mu=0$.
Quark-number susceptibilities have been studied by
S.Gottlieb {\it et al}~\cite{go} and  give us the signals for the chiral-
symmetry-restoration phase transition. Study of response of hadron masses
to chemical potential is now under investigation by $QCD-TARO$ 
 collaboration~\cite{ta}.

Another way is to study  LQCD for color SU(2) system where
action remains real.
 Hamiltonian approach at zero temperature and finite
density has been studied by E.B.Gregory {\it et al}~\cite{gr}, and gives us
some
results for vacuum energy, chiral condensate, baryon-number density and its
susceptibility.
Numerical simulation at non zero chemical potential
  has been performed by S.Hands
{\it et al}~\cite{ha}, by using several algorithms;
 HMD  algorithm ; HMC  algorithm ; 
 Two-Step Multi-Boson ( TSMB ) algorithm~\cite{mo}
on rather small lattices ( mostly on $6^4$ lattice ) at zero temperature.
They studied relation of chiral condensate, pion mass and chemical potential
at $\mu  > \frac{m_{\pi}}{2}$ as well as $\mu  < \frac{m_{\pi}}{2}$.

In this paper, we focus on
 the effects of chemical potential to masses and condensate at
finite temperature
( $\mu$=0.02, 0.05, 0.1 and 0.2 ) on larger lattices $12^{3}\times 4$ and 
 $12^2\times24\times4$.
One of interests is precise study of masses of meson with nonzero
chemical potential since QCD Sum Rule suggests linear dependence on density.
\cite{hl}
Firstly, we shall observe chiral condensate with nonzero
chemical potential near critical temperature.
Secondly, we shall examine meson masses consisting of
light-light quarks and light-heavy quarks.
 The organization of the paper is as follows: In Section 2 we state 
our procedure.  Simulation is described in section 3.
We show some preliminary results of the present simulation in Section 4.
Summary and conclusions are stated in Section 5.

%***********************end  section 1 *********************
%***********************start  section 2 *********************
\section{STAGGERED FERMION ACTION AT NONZERO CHEMICAL POTENTIAL}

At non zero chemical potential, staggered fermion matrix  ${M}$ is
given by
\begin{equation}
\begin{array}{rcl}
{M_{x,y}(U,\mu)}&=&
2m\delta_{xy}+\sum\limits_{\nu=1}^{3}\eta_{\nu}(x)[U_{x,{\nu}}
\delta_{y,x+\hat{\nu}}-U^{\dagger}_{y,\nu}\delta_{y,x-\hat{\nu}}]\\
&&{}+ \eta_{4}(x)[e^{\mu}U_{x,{4}}
\delta_{y,x+\hat{4}}-e^{-\mu}U^{\dagger}_{y,4}\delta_{y,x-\hat{4}}],\\
& & \\
\end{array}
\end{equation}
where ${U}$ is  SU(2) link variables
and $\mu$ is chemical potential. $\eta_{\nu}$ is staggered phase and
$m$ is quark mass parameter.

The matrix $M$ satisfies the following relation

\begin{equation}
\begin{array}{rcl}
{M_{x,y}(U,\mu)}&=&
(-1)^{sig(x)-sig(y)}{M^{\dagger}_{x,y}(U,-\mu)}.\\
& & \\
\end{array}
\end{equation}

The path integral for the partition function is given by 
\begin{equation}
\begin{array}{rcl}
{Z}& = &
\int[DU]det(M)^{N_f/4}e^{-S_g},\\
& & \\
\end{array}
\end{equation}
where $S_{g}$ is plaquette action written as ,
\begin{equation}
\begin{array}{rcl}
{S_{g}}& = &
\beta\sum\limits_{P}[1-\frac{1}{4}Tr(U_{P}+U_{P}^{\dagger})],\\
& & \\
\end{array}
\end{equation}
where $\beta= \frac{4}{g^{2}}$ .

In cases of simulations by HMC algorithm,
we need positive definiteness of hamiltonian.
Since $\det(M)$ is real for SU(2) system,
the positive definiteness condition is achieved by
taking $\det(M)^2$ which includes $n_f=8$ dynamical fermions.
Note that the standard even-odd partitioning
of the fermion matrix which is commonly used to reduce the number of flavors
can not be applied for simulations at finite chemical potential.
Thus $n_f=8$ is the minimum number of flavors for
$SU(2)$ HMC simulations with dynamical fermions.
The present study is performed with $n_f=8$.

%-----------------end  subsection 2.1---------------------
%***********************start  section 3 *********************
\section{SIMULATIONS}

Firstly, we have done simulation on $6^4$ lattice and confirmed that
our results are consistent with the previous work by S.Hands 
 {\it et al}~\cite{ha}.
After that, numerical simulations on $12^3\times 4$ and
$12^2\times24\times 4$ lattices
have been carried out for parameters summarized in Table 1.

%%%%%%%%%%%%%%%%%%% table %%%%%%%%%%%%%%%%%%%%%%%%%%%%%%
\begin{center}
\begin{table}[htbp]
\caption{List of the data sample}
\vspace{0.2cm}
\begin{center}
\begin{tabular}{|c|c|c|c|c|c|l|}
\hline
% flavor number & quark mass & lattice size & quark mass & $\beta$\\
$N_{F}$ & $m_{q}$ & $\mu$ & lattice size & $\beta$\\
\hline
{} & 0.05 & 0.0 , 0.02 & $6 \times 6 \times 6 \times 6$ &
 0.1$\sim$2.5 ( for test )\\
\cline{4-5}
%\hline
8 & 0.07 & 0.05 , 0.1 &  $12 \times 12 \times 12 \times 4$ &
 1.0$\sim$2.0( to determine $\beta _{c}$ )\\
\cline{4-5}
%\hline
{} & 0.1 & 0.2  & $12 \times 12 \times 24 \times 4$  &
 1.1 ( for simulations )\\
\hline
\end{tabular}
\end{center}
\end{table}
\end{center}

Number of configuration for every
parameter is also given in Table 2.

\begin{center}
\begin{table}[htbp]
\caption{Statistics of the data}
\vspace{0.2cm}
\begin{center}
\begin{tabular}{|c|c|c|c|c|c|l|}
\hline
{} & $\mu=0.0$ & $\mu=0.02$ & $\mu=0.005$ & $\mu=0.1$ & $\mu=0.2$\\
\hline
$m_q=0.05$ & 120 & 180 & 160 & 80 & 30\\
\hline
$m_q=0.07$ & 140 & 170 & 160 & 160 & 30\\
\hline
$m_q=0.10$ & 180 & 170 & 160 & 140 & 70\\
\hline
\end{tabular}
\end{center}
\end{table}
\end{center}

Most of the simulations are carried out on small chemical potential
%$\mu < \frac{m_{\pi}}{2}$
 and at fixed $\beta =1.1$ which is little  below $\beta_{c}$.
Time step of HMC algorithm is dt=0.01.
After 1000 trajectories of thermalization, data are taken at every
100 trajectories.
%----------------------------------------------------------
Since our simulation is restricted in a region of small chemical potential,
the algorithm works without difficulty.  But, necessary CPU time
increases with the chemical potential. A typical trend is shown
in table 3.
At $\mu=0.2$, average CPU time is 3.5 times longer than that at $\mu=0.0$.
\begin{center}
\begin{table}[htbp]
\caption{CPU time of getting one configuration
  at $\beta=1.1$, $m_q=0.05$}
\vspace{0.2cm}
\begin{center}
\begin{tabular}{|c|c|c|c|c|c|l|}
\hline
{} & 0.0 & 0.02 & 0.05 & 0.1 & 0.2\\
\hline
time(second) & 2223.3 & 2305.7 & 2536.0 & 3122.5 & 7568.0\\
\hline
\end{tabular}
\end{center}
\end{table}
\end{center}

%%%%%%%%%%%%%%%%%%%%%%%%%%%%%%%%%%%%%%%%%%%%%%%%%%%%%%%%%%%%
%----------------fig.1 and fig.2-------------------------------
\begin{figure}[htbp]
\epsfig{figure=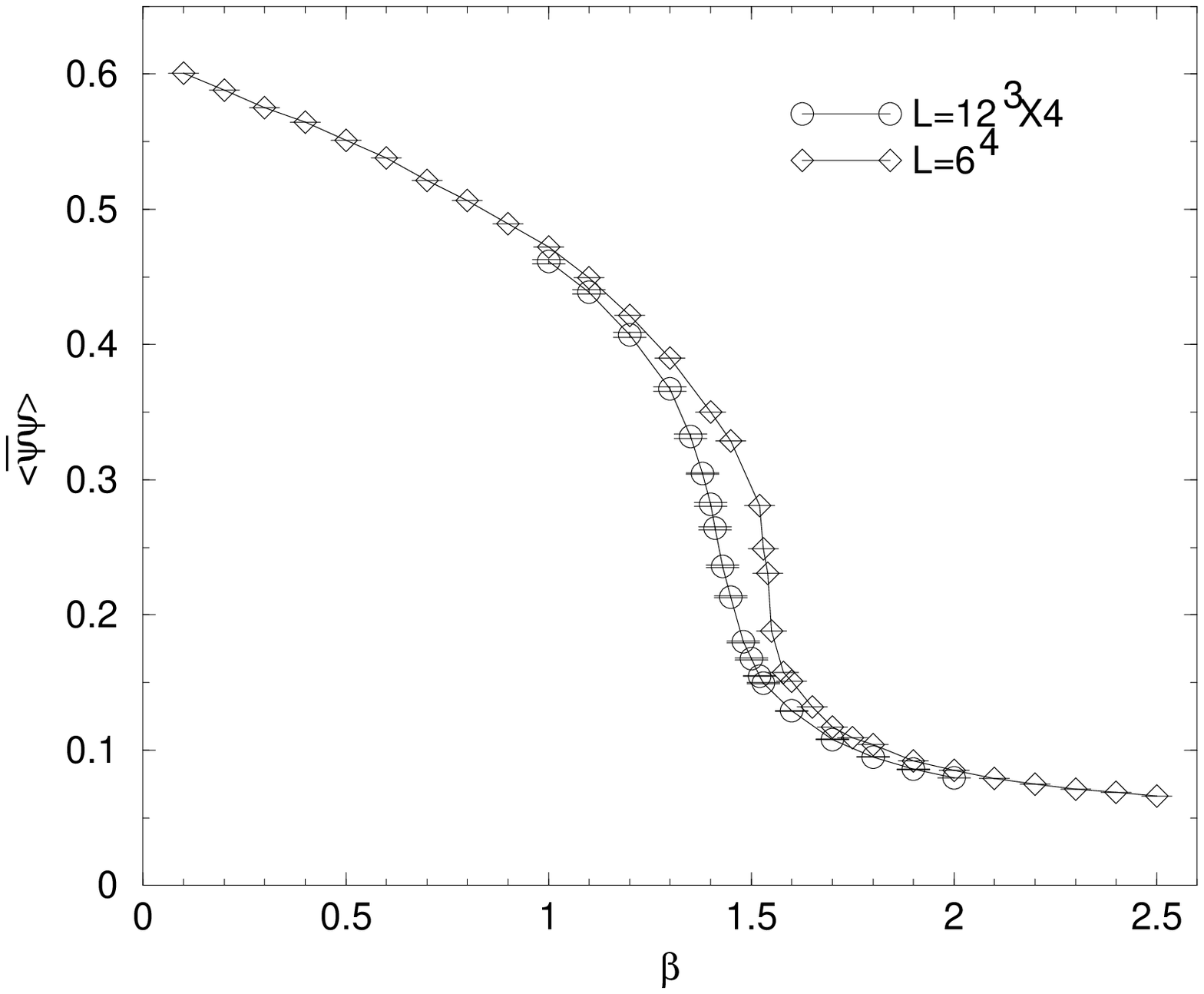,height=1.8in}
\hfill
\epsfig{figure=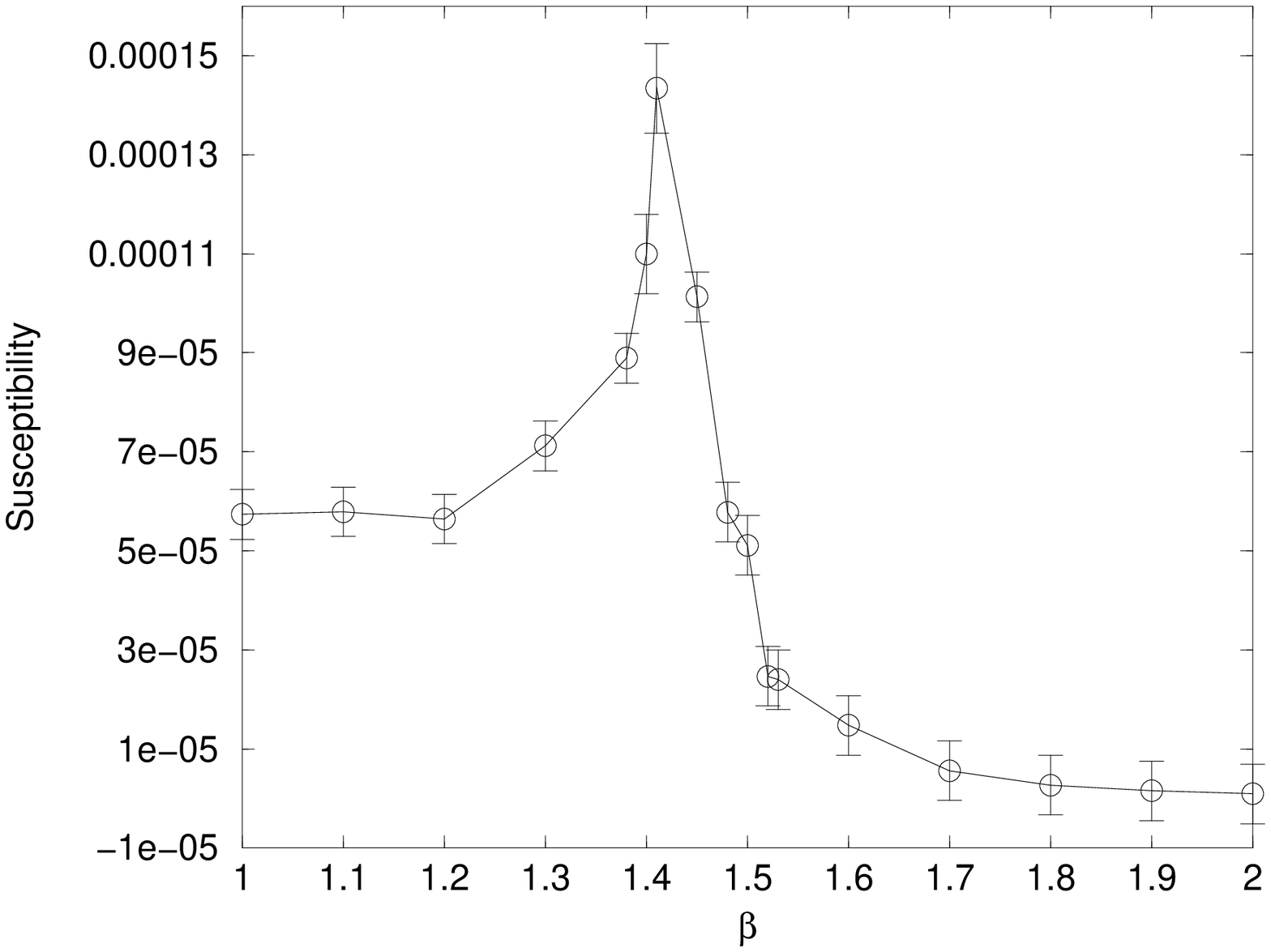,height=1.8in}

\hspace{28mm}(a)\hspace{58mm}  (b)
\caption{(a)  $<\bar{\psi}\psi>$ vs.\hspace{1mm} $\beta$ for
$L=6^{4}$(diamonds) and
$L=12^{3}\times4$(circles),\hspace{28mm}
(b)   Susceptibility of $<\bar{\psi}\psi>$ vs.\hspace{1mm} $\beta$
for $L=12^{3}\times4$.}
\end{figure}
%----------------------------------------------------------------

\section{RESULTS}
Firstly we determine phase transition point  $\beta_{c}$
at $\mu=0$ on $12^3\times 4$ lattice.

Chiral condensate
$<\bar{\psi}\psi>$ versus $\beta$ is shown in figure 1(a).  Chiral
restoration occurs at around $\beta\approx 1.4$. In order to determine
the critical point more precisely, susceptibility of the condensate
is analyzed as shown in figure 1(b).  By this analysis, the critical
point for $m=0.07$ is determined as

\begin{equation}
\beta_c =1.41 \pm 0.03
\end{equation}

This value is slightly smaller than that of finite size crossover transition
point $\beta_c=1.54$ on $6^4$ lattice.
%%%%%%%%%%%%%%%%%%%%%%%%%%%%%%%%%%%%%%%%%%%%%%%%%%%%%%%%%%%%%%%%%%%%

Nextly, we study quark condensate and pseudoscalar meson mass at
small but finite $\mu$.  We stay in confinement phase ($\beta=1.1$).
In figure 2(a), we plot  the   $<\bar{\psi}{\psi}>$ as a function 
 of  $\mu$ ( $0.0 \sim 0.2$ ) at several  $m$ (0.05,0.07,0.1).
As shown in the figure, the condensate decreases with the chemical
potential. But it varies very weakly around $\mu=0$.

Pseudoscalar meson mass is extracted in a standard way.  Typical examples
of pseudoscalar correlator is shown in figure 2(b).  In the present region
of study, single pole fit gives reasonable fitting and results for
pion mass squared are  shown in figure 3(a).  In order to examine the influence
of chemical potential on hadron mass, we examine pseudoscalar meson
consisting of light and heavier quarks.  figure 3(b) shows
their masses at small $\mu$.  As shown in the figure, they are
very stable and no appreciable dependence on chemical potential
is seen in this region.

%%%%%%%%%%%%%%%%%%%%%%%%%%%%%%%%%%%%%%%%%%%%%%%%%%%%%%%%%%%%%%%%%%%%
\begin{figure}[htbp]
\epsfig{figure=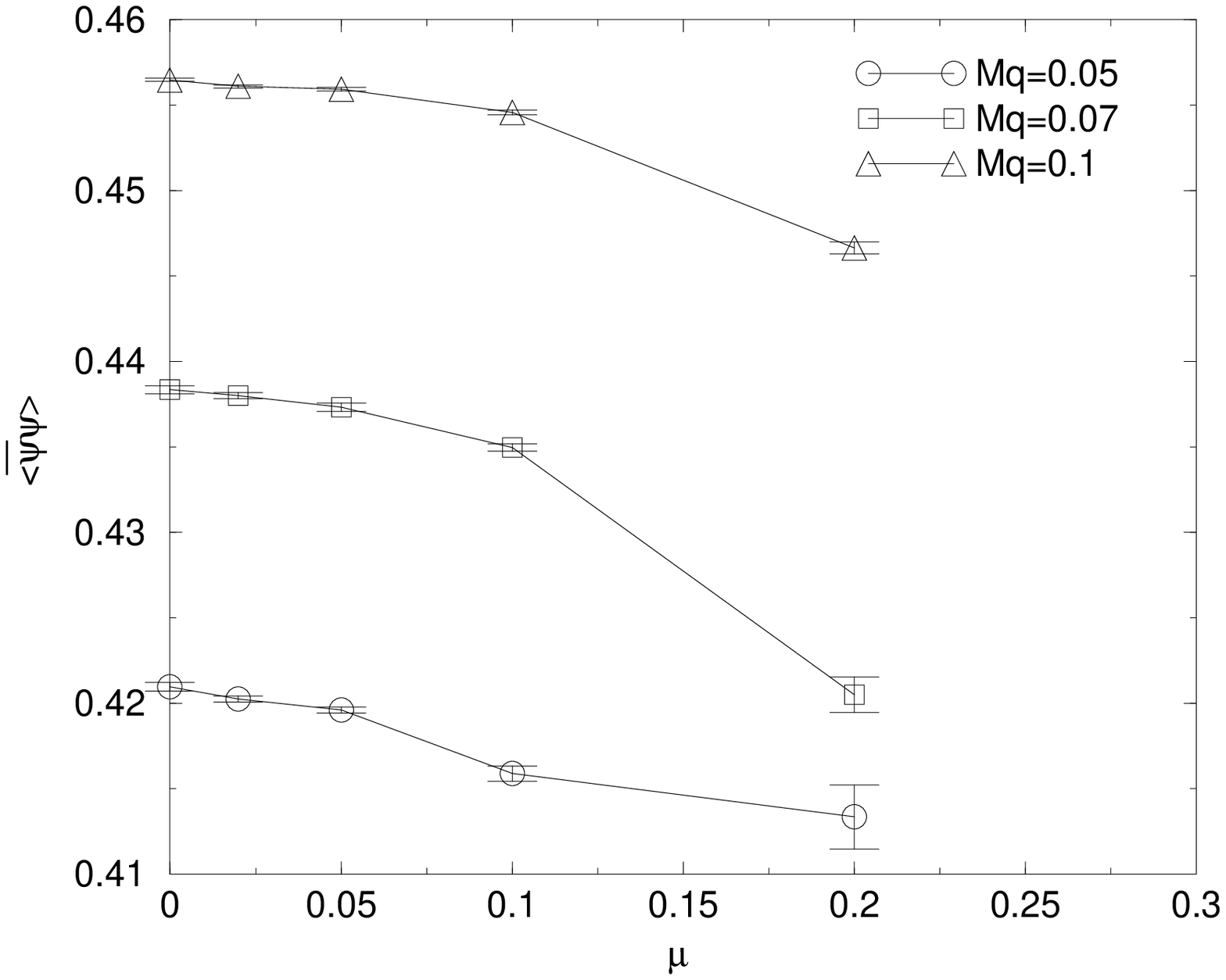,height=1.8in}
\hfill
\epsfig{figure=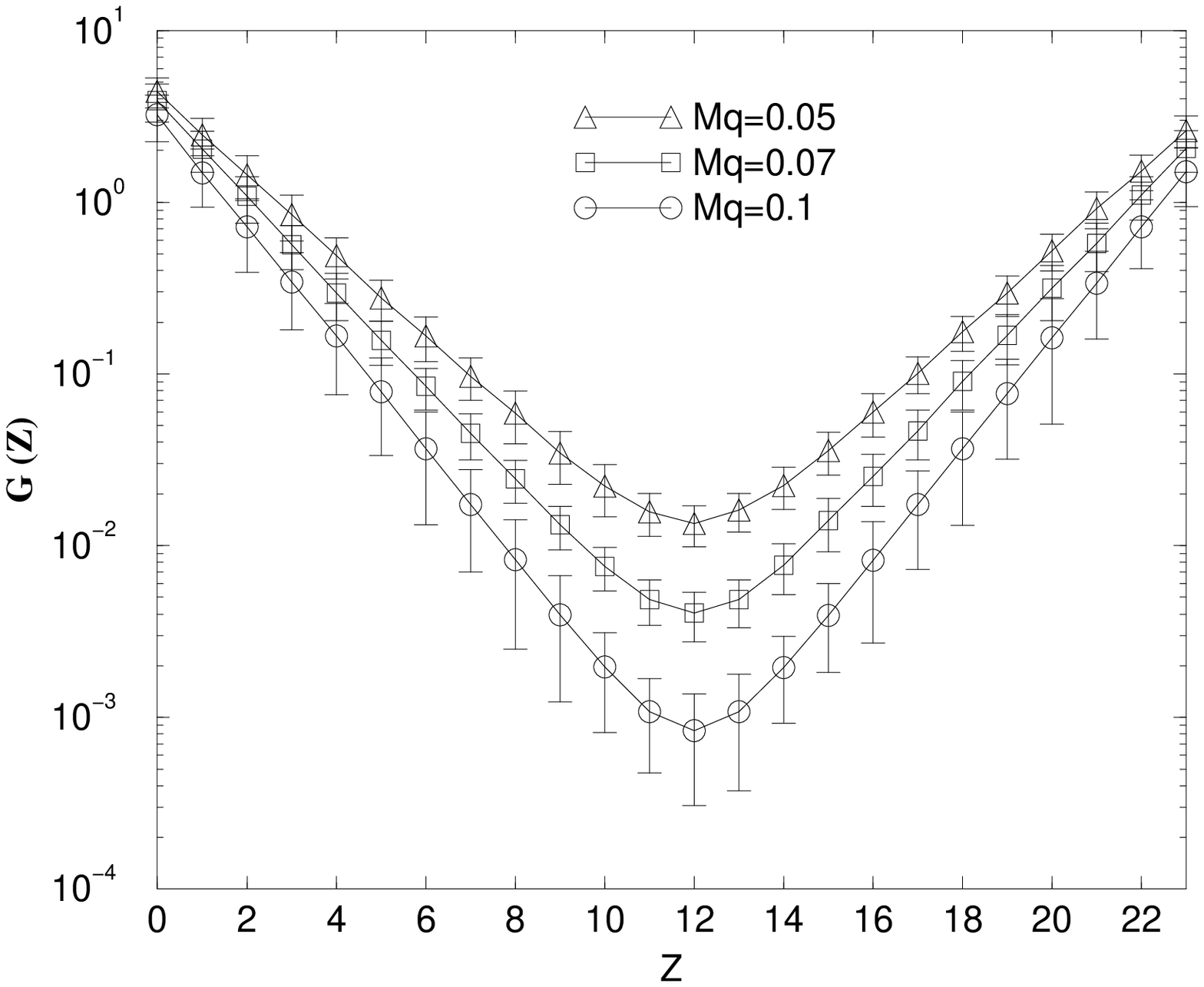,height=1.8in}

\hspace{28mm}(a)\hspace{58mm}  (b)
\caption{(a)  $<\bar{\psi}\psi>$ vs.\hspace{1mm}$\mu$  for $m_q=0.05, 0.07,
0.1$;
\hspace{1mm} $\beta=1.1$.\hspace{100mm}
(b)  correlation function   of $<\bar{q}q>$ vs.\hspace{1mm}$\beta$
for $\beta=1.1$.
\hspace{1mm} $\mu=0.1$, $m_q=0.05, 0.07, 0.1$.}
\end{figure}
\begin{figure}[ht]
\epsfig{figure=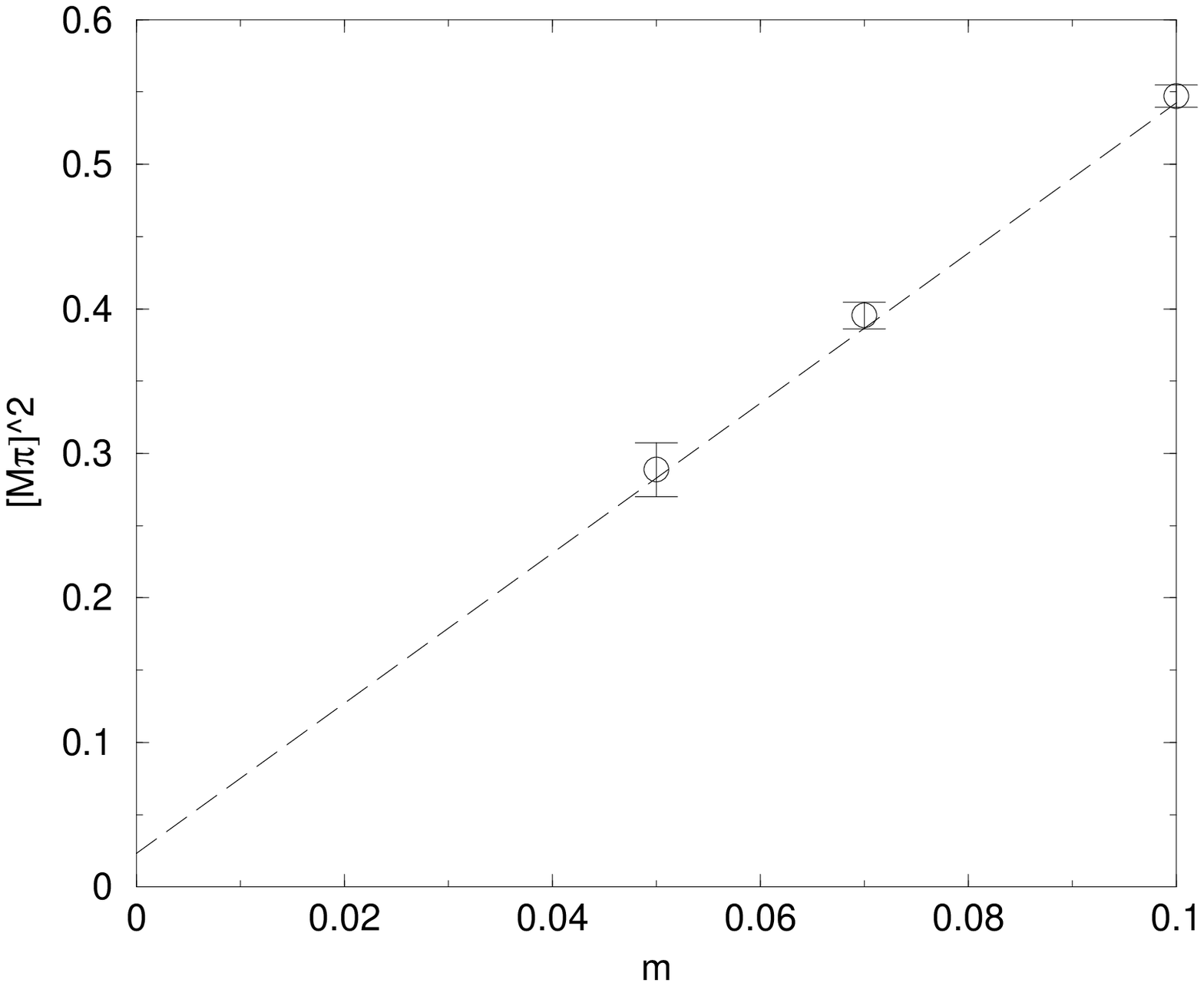,height=1.8in}
\hfill
\epsfig{figure=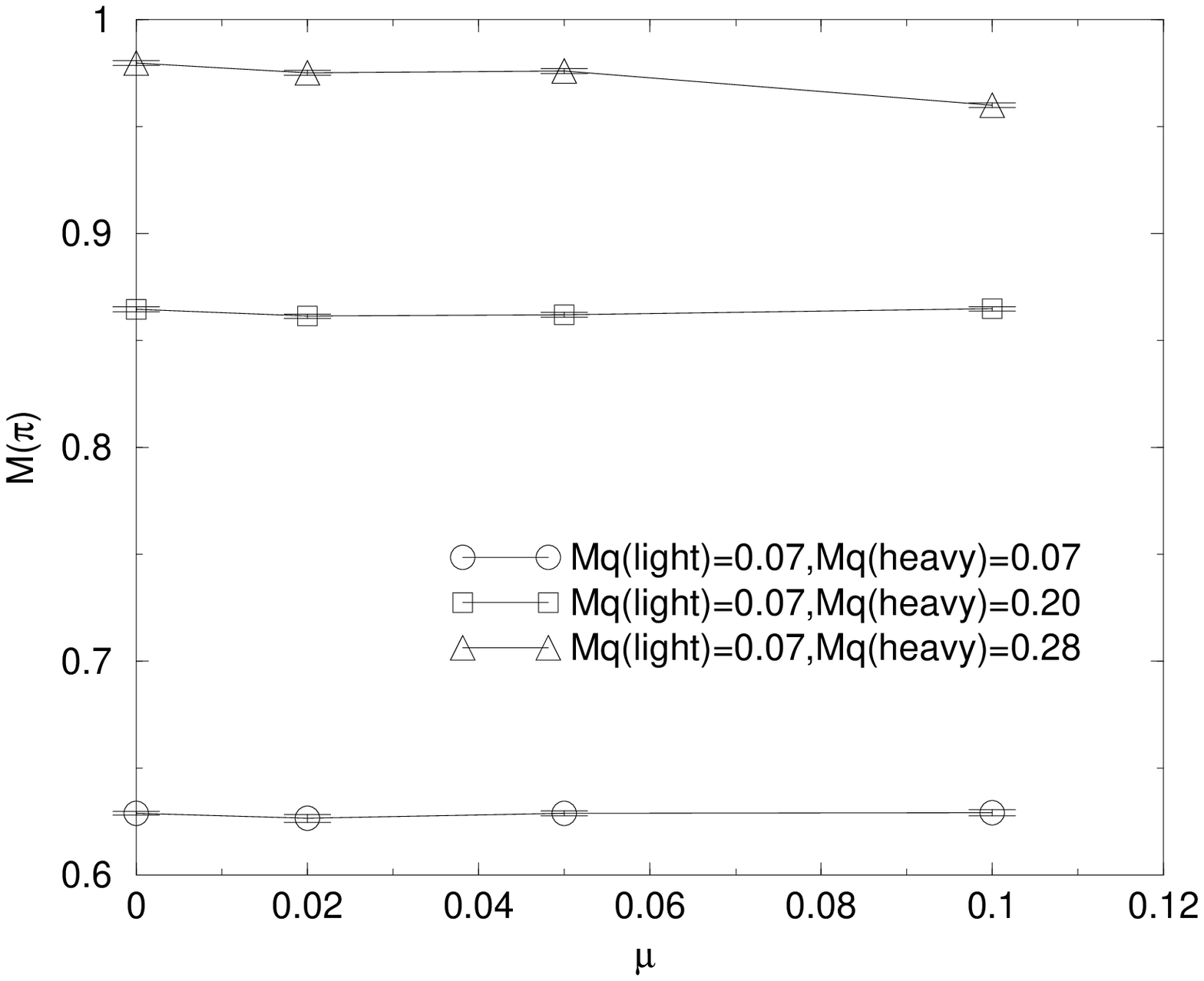,height=1.8in}

\hspace{28mm}(a)\hspace{58mm}  (b)
\caption{(a)  ${M_{\pi}}^{2}$ vs. $m_q$ for $\beta=1.1$, $\mu=0$
;
(b)  ${M_{\pi}}$ vs. $\mu$
for $\beta=1.1$,
}
\end{figure}
%%%%%%%%%%%%%%%%%% end  Sec. 4 %%%%%%%%%%%%%%%%%%%%%%%%%%%%%%%%%

%%%%%%%%%%%%%%%%%% start  Sec. 5 %%%%%%%%%%%%%%%%%%%%%%%%%%%%%%%%%
\section{SUMMARY AND CONCLUSIONS}
In this paper, we study  8 flavors SU(2) QCD
at finite temperature and small $\mu$ on $12^3\times 4$ and
$12^2\times24\times 4$ lattices.
The chiral transition at $\mu=0$ is determined to be $\beta_c=1.41$.
Chiral condensate decreases with $\mu$ but variation is very weak
around $\mu=0$. Mass of pseudoscalar meson consisting of light
and heavier quarks is stable against chemical potential.
%%%%%%%%%%%%%%%%%% end Sec. 5 %%%%%%%%%%%%%%%%%%%%%%%%%%%%%%%%%

\end{document}